\documentclass[11pt,a4paper]{article}
\usepackage{times}
\usepackage[height=19cm,width=13cm]{geometry}

\begin{document}
\vskip 5cm
\title{ Dissipative Two-Fluid  Models}
\author{\\ Henri Gouin \  and \ Sergey Gavrilyuk\\  \\
{ {\ Universit\'e d'Aix-Marseille, 13397 Marseille Cedex 20\ \ France}}%
\\ (\small E-mails:\ henri.gouin@univ.u-3mrs.fr;
sergey.gavrilyuk@univ.u-3mrs.fr)\\ \\ }
\date{{\small {{ To Guy Boillat with friendship}}}}
\maketitle

\vskip 1cm
\begin{abstract}
From Hamilton's principle of stationary action, we derive
governing equations of  two-fluid mixtures and extend the model to
the dissipative case without chemical reactions. For both
conservative and dissipative cases, an algebraic identity
connecting equations of momentum,  mass, energy  and entropy  is
obtained by extending the Gibbs identity in dynamics. The obtained
system  is hyperbolic for small relative velocity of the phases.
\end{abstract}

\vskip 1cm

\section{Introduction}

The knowledge of governing equations for fluid mixtures is scientifically
and industrially an important challenge. Many authors derived the governing
system by using axioms of balance of mass, momentum, energy and second law
of thermodynamics \cite{Bowen}. The mixtures were considered as a collection
of different media co-existing in the physical space. For example, the
balance law of momentum is given in the form :
\begin{equation}
\frac{\partial }{\partial t}\int_{D}\rho _{\alpha }\,{\mathbf{u}}_{\alpha
}\,dD\ +\int_{\partial D}\rho _{\alpha }\,{\mathbf{u}}_{\alpha }\,{u}%
_{\alpha n}\ d\sigma =\int_{\partial D}T_{\alpha }\ {\mathbf{n}}\ d\sigma \
+\int_{D}{\mathbf{b}}_{\alpha }\ dD,\ \ \alpha \ =\ 1,2  \label{(balance)}
\end{equation}%
\noindent where $\,D\,$ is a fixed volume, $\,\partial D\,$ is its
boundary, $\, {\mathbf{n}}\,$ is the unit normal to $\,\partial D$,
$\,\rho _{\alpha }$ are the densities of components,
${\mathbf{u}_{\alpha }}$ are the associated velocities, $\,u_{\alpha
n}$ are the normal part of the velocities at $ \,\partial
D,\,T_{\alpha }\,$ are the stress tensors and ${\mathbf{b}}_{\alpha
}$ are the volume forces associated with internal forces and
interaction between components. The principle of material
frame-indifference requires that $\ T_{\alpha }\ $ and $\
{\mathbf{b}}_{\alpha }\ $ depend on the thermodynamic parameters of
the mixture and on the relative velocity $\
{\mathbf{w}}={\mathbf{u}}_{2}-{\mathbf{u}}_{1}$. To include the
added mass effect into consideration, one should be supposed that
$\,T_{\alpha }\,$ and ${\mathbf{b}}_{\alpha }\,$ depend also on
accelerations of phases. The structure of this dependence which is
an important source of interaction, is not clear: should it be
frame-indifferent or simply Galilean invariant? The method of
balance laws (\ref{(balance)}) does not give a definite answer to
this question.\newline It exists a different approach based on
Hamilton's principle \cite{Bedford}-\cite{Gouin} which is used for
construction of conservative (non-dissipative) mathematical models
of continuous media with complex internal structure. The terms
including interaction between different components of the mixtures
do not require constitutive postulates difficult to interpret
experimentally. They come from the direct knowledge of a unique
potential for the mixture. According to \cite{Nigmatulin}, we call a
\emph{homogeneous mixture} if each component of the mixture occupies
the whole volume of the physical space, and a \emph{heterogeneous
mixture} if each component occupies only a part of the mixture
volume. In this paper, we consider only homogeneous binary mixtures,
but the method can be extended to the case of heterogeneous mixtures
\cite{Saurel1}.  The plan for the article is as follows :

In section 2, we formulate an extended form of Hamilton's principle
of stationary action allowing us to obtain the governing equations
of motion. The Lagrangian is the difference between the kinetic
energy depending on the reference frame and a thermodynamic
potential which is a Galilean invariant. The equations of motion
introduce two new vector fields different from the velocity fields
by taking into account the relative velocity of the components. They
play the same role as the velocity field does in the case of a
single fluid.

In section 3, the governing equations are extended to the
dissipative case without chemical reactions. An algebraic identity
connecting equations of momentum, equations of mass, energy equation
and equations of entropy is obtained. This identity can be
considered as the dynamic form of the Gibbs identity.

In section 4, we justify the compatibility of the governing equations with
the second law of thermodynamics.

In section 5, Fick's law is derived. We prove that Fick's law is not yet a
linear phenomenological law but a direct consequence of governing equations
and Stokes drag force hypothesis.

In Section 6, we check the properties of hyperbolicity of the
governing system  for small relative velocity of phases.

As a convention, in the following we shall use asterisk $\
^{\prime\prime}\star ^{\prime\prime}\ $ to denote \textit{conjugate}
mappings or \textit{covectors} (vector lines); subscripts $\ \alpha\ =\ 1,2\
$ indicate the parameters of the $\, \alpha^{th}$ component;  the symbol $I$
indicates the identity; $\ \nabla $ means the gradient operator-line; $\
\nabla^{\star} $ means the gradient operator-column; $\ {\mathbf{a}}
^{\star}\ {\mathbf{b}} $ means the \textit{scalar product} of vectors $\ {%
\mathbf{a}},\ {\mathbf{b}}\ $  (the vector line is multiplied by the vector
column); $\ {\mathbf{a}}\ {\mathbf{b}}^{\star} $ means the \textit{tensor
product} of vectors $\ {\mathbf{a}},\ {\mathbf{b}}\ $ (the vector column is
multiplied by the vector line); $\ A\ {\mathbf{a}} $ means the product of
the mapping $\ A\ $ by a vector $\ {\mathbf{a}}\ $; \ ${\mathbf{b}}^{\star}\
A $ means the covector $\ {\mathbf{c}}^{\star}\ $ defined by the rule $\ {%
\mathbf{c}}^{\star}\ =\ (A^{\star}\ {\mathbf{b}})^{\star}$; $div\ A $
denotes the divergence of a linear transformation $\ A\ $ which is a
covector defined as follows: for any vector $\ {\mathbf{a}}$,
\[
div\ (A\ {\mathbf{a}})\ = \ (div A)\ {\mathbf{a}}\ +\ tr\ \bigg(A\ \frac{%
\partial{\mathbf{a}}}{ \partial{\mathbf{x}}}\bigg).
\]

\section{Governing equations in conservative case}

In paper \cite{Gavrilyuk2}, we considered a pure \emph{mechanical}
case (without entropy). Now we consider the general case. We take
the Lagrangian of the binary system in the following form:
\begin{equation}
L \ =\ \sum^{2}_{\alpha=1}\ {\frac{1}{2}}\ \rho_\alpha\ {\mathbf{u}}%
_\alpha^2\  - \rho_\alpha\Omega_\alpha \, - \, W\ (\rho_1, \rho_2
, s_1 , s_2 , {\mathbf{w}}) \label{Lagrangien}
\end{equation}
where in the whole paper  the summation  is over fluid components
$(\alpha =1,\,2)$ and $\,\rho _{\alpha }\,$ are the densities of
components, $\,s_{\alpha }\,$ are the specific entropies,
$\,{\mathbf{u}}_{\alpha }\,$ are the velocities, $\ {\mathbf{w}}\,
=\, {\mathbf{u}}_2-{\mathbf{u}}_1\ $ is the relative velocity, $\
\Omega_\alpha\ $ are the external force potentials, $\ W $ is a
potential per unit volume of the mixture. The dependance of $\ W $
with respect to the relative velocity is analog to take into account
the added mass effect in heterogeneous two-fluid theory as it was
done by Geurst \cite{Geurst}.  The fact that $W$ depends on two
entropies is classically adopted in the literature
\cite{Bedford,Gouin,Muller,Muller2}. The potential $\ W $ is related
with the internal energy $\ U\; $ of the mixture through the
transformation
\begin{equation}
U\ =\ W -\ \frac{\partial W }{ \partial{\mathbf{w}}}\ {\mathbf{w}}
\label{(internalenergy)}
\end{equation}
so that the total energy of the system is \cite{Gavrilyuk2}:
\[
\varepsilon \ = \ \sum^{2}_{\alpha=1}\ {\frac{1}{2}}\ \rho_\alpha {\mathbf{u}%
}_\alpha^2\, + \rho_\alpha\Omega_\alpha +\, U
\]
Let us note that to define the internal energy of one-velocity media
it is useful to consider a moving coordinate system in which the
elementary volume of the continuum is at rest. The total energy of
the continuum with respect to this system is called the internal
energy of the medium. For a two-velocity medium, there is no
reference frame in which any motion could be disregarded. This is
the reason why the standard definition of internal  energy is
dependent on the relative motion of components. The formula
(\ref{(internalenergy)}) implies that the internal energy $\ U\ $ is
a Galilean invariant. The dependence of $\ U\ $ (or $\ W $) on $\
{\mathbf{w}}\ $ is an important property of multicomponent fluid
mixtures. \newline
Let $\ {\mathbf{x}}\ $ be the Eulerian coordinates, $\ t\ $ be the time, $\ {%
\mathbf{X}}_\alpha \, $ be the Lagrangian coordinates of each
component. The mass and the entropy conservation laws in the
Eulerian coordinates are:
\begin{equation}
\frac{\partial\rho_\alpha}{\partial t}\ + \ div\ (\rho_\alpha\ {\mathbf{u}}%
_\alpha)\ = \ 0\ \,,\ \ \ \frac{\partial}{\partial t} (\rho_\alpha\,
s_\alpha)\ + div\ (\rho_\alpha s_\alpha\, {\mathbf{u}}_\alpha)\ = \
0\ . \label{(mass)}
\end{equation}
In the Lagrangian coordinates, equations (\ref{(mass)}) are equivalent to:
\begin{equation}
\rho_\alpha\ \det\ F_\alpha\ =\ \rho_{\alpha 0}\ ({\mathbf{X}}_\alpha) \,,\
\ \ s_\alpha\ =\ s_{\alpha 0}\ ({\mathbf{X}}_\alpha),  \label{(determinant)}
\end{equation}
where
\begin{equation}
F_\alpha\ = \frac {\partial{\mathbf{x}}}{ \partial{\mathbf{X}}_\alpha}
\label{(Jacobian)}
\end{equation}
is the deformation gradient at $\, {\mathbf{X}}_\alpha ; $  $\, \rho_{\alpha
0}\, ({\mathbf{X}}_\alpha)$ and $s_{\alpha 0}\, ({\mathbf{X}}_\alpha)\, $ do
not depend on $t$.  The relation between the Eulerian and Lagrangian
coordinates is given by the local diffeomorphism $\ {\mathbf{x}} = {\mathbf{%
\phi}}_\alpha\ ({\mathbf{X}}_\alpha, t),\ $ where ${\mathbf{\phi}}_\alpha\ ({%
\mathbf{X}}_\alpha , t)\ $ is the solution of the Cauchy problem :
\[
{\frac{d{\mathbf{\phi}}_\alpha}{dt}} \ = \ {\mathbf{u}}\ ({\mathbf{\phi}}%
_\alpha, t)\, ,\ \ \  {\mathbf{\phi}}_\alpha\ ({\mathbf{X}}_\alpha,
0)\ = \ {\mathbf{X}}_\alpha
\]
Let $\ {\mathbf{X}}_\alpha\ = {\mathbf{\psi}}_\alpha\ ({\mathbf{x}} , t) \ $
be its inverse mapping ($\ {\mathbf{\phi}}_\alpha\ \circ\ {\mathbf{\psi}}%
_\alpha\ =\ I$). We define the virtual motion of the mixture such that \cite%
{Gouin,Serrin}:
\[
{\mathbf{x}}\ =\ {\mathbf{\Phi}}_\alpha\ ({\mathbf{X}}_\alpha, t,
\varepsilon_\alpha)\ \ \ ,\ \ \ {\mathbf{X}}_\alpha\ =\ {\mathbf{\Psi}}%
_\alpha\ ({\mathbf{x}}, t , \varepsilon_\alpha)\ \ \ ,\ \ \ {\mathbf{\Phi}}%
_\alpha\ \circ\ {\mathbf{\Psi}}_\alpha\ \ =\ I\
\]
\[
{\mathbf{\Phi}}_\alpha\ ({\mathbf{X}}_\alpha , t , 0)\ =\ {\mathbf{\phi}}%
_\alpha\ ({\mathbf{X}}_\alpha , t)\ \ \ ,\ \ \ {\mathbf{\Psi}}_\alpha\ ({%
\mathbf{x}} , t , 0)\ =\ {\mathbf{\psi}}_\alpha\ ({\mathbf{x}} , t),
\]
where $\, \varepsilon_\alpha $ belong to a vicinity of zero.
\noindent The Lagrangian and Eulerian virtual displacements are
defined respectively as :
\begin{equation}
\delta {\mathbf{X}}_\alpha\ = \ {\frac{\partial\ {\mathbf{\Psi}}_\alpha}{%
\partial\varepsilon_\alpha}}\ ({\mathbf{x}} , t , \varepsilon_\alpha)\
\vert\ _{\varepsilon_\alpha=0}\ , \ \ \ \delta_\alpha {\mathbf{x}}\ = \ {%
\frac{\partial{\mathbf{\Phi}}_\alpha}{\partial\varepsilon_\alpha}}\ ({%
\mathbf{X}_\alpha} , t , \varepsilon_\alpha)\ \vert\ _{\varepsilon_\alpha=0}
\ .  \label{(displacements)}
\end{equation}
The definitions (\ref{(Jacobian)}-\ref{(displacements)}) imply the following
relation between $\ \delta{\mathbf{X}}_\alpha\ $ and $\ \delta_\alpha{%
\mathbf{x}}\ $ \cite{Gouin}:
\begin{equation}
\delta_\alpha {\mathbf{x}}\ = -\ F_\alpha\ \delta {\mathbf{X}}_\alpha
\label{(displacement)}
\end{equation}
The variations of $\ {\mathbf{u}}_\alpha\, (t , {\mathbf{x}})  , \
\rho_\alpha\, (t, {\mathbf{x}})\ $ and $\, s_\alpha\, (t ,
{\mathbf{x}})\ $ are deduced from (\ref{(Jacobian)}) -
(\ref{(displacements)}) and from the definition of the Lagrangian
coordinates $\ {\mathbf{X}}_\alpha\ $:
\[
{\frac{d_\alpha {\mathbf{X}}_\alpha}{dt}}\ = \ 0\ ,\ \ \ {\frac{d_\alpha}{dt}%
}\ = \ {\frac{\partial}{\partial t}}\ + \ {\mathbf{u}}^{\star}_\alpha\
\nabla^{\star}.
\]
We obtain in Appendix A the values of $\ \delta {\mathbf{u}}_\alpha\, ({%
\mathbf{x}} , t)$, $\delta \rho_\alpha\ ({\mathbf{x}} , t)$ and $\delta
s_\alpha\ ({\mathbf{x}} , t)$, where $\delta f(t,{\mathbf{x}})$ means the
variation of $f$ when $t,\,{\mathbf{x}}$ are fixed and $div_\alpha (\delta{%
\mathbf{X}}_\alpha ) $ means the divergence with respect to the coordinates $%
{\mathbf{X}_\alpha}$. We note that in \cite{Gavrilyuk2} we used
different but equivalent expressions for these variations. Using the
definition (2) of the Lagrangian $\ L \,$ as a function of
$\rho_\alpha, {\mathbf{u}}_\alpha, s_\alpha$, we introduce the
following quantities:
\begin{equation}
\cases{R_\alpha\ \equiv\ \displaystyle \frac{\partial L}{
\partial\rho_\alpha}\ = \frac {1}{ 2}\  {\mathbf u}_\alpha^2 \,
- \frac {\partial W}{ \partial\rho_\alpha}\ -\ \Omega_\alpha,\cr\cr\cr
{\mathbf K}_\alpha^\star\ \equiv\ \displaystyle \frac {1}{ \rho_\alpha}\
\frac {\partial L}{ \partial{\mathbf u}_\alpha}\ = \ {\mathbf
u}^\star_\alpha\ - \frac {(-1)^\alpha }{ \rho_\alpha}\ \frac {\partial W }{
\partial {\mathbf w}} ,\cr\cr\cr \rho_\alpha\ \theta_\alpha\ \equiv\
\displaystyle -\frac{\partial L}{ \partial s_\alpha}\ = \frac {\partial W }{
\partial s_\alpha }}  \label{(definitions)}
\end{equation}
We note that a best set of independent variables is :
$\rho_\alpha,\, {\mathbf{j}}_\alpha=\rho_\alpha
{\mathbf{u}}_\alpha,\, \rho_\alpha s_\alpha$. However, in this case,
the corresponding derivation should be given in four-dimensional
space \cite{Gavrilyuk4}. For the sake of simplicity we use the first
set of independent variables. The last formula defines the
thermodynamic temperature $\, \theta_\alpha\, $ of each component
which is now a dynamical quantity depending on the relative velocity
of the components.\newline Let $\ \omega\ = \ {D}\ \times\ [t_1\ ,
t_2]\ $ be the domain in the four-dimensional space $\,
({\mathbf{x}} , t)\, $ and $\, \omega_\alpha\, $ be its image in the
$({\mathbf{X}}_\alpha, t) $-space. Here $\ [t_1, \ t_2]\ $ is a time
interval and $\, {\ D}\, $ is a fixed domain. We consider Hamilton's
principle in the form,
\[
\delta_\alpha a \equiv \ \delta_\alpha\int_\omega\ L\ d\omega\ = \ 0
\]
under constraints (\ref{(mass)}) where $\ \delta_\alpha a\ $ are variations
of \, $a$\, associated with the variation of ${\mathbf{X}}_\alpha = {\mathbf{%
\Psi}}_\alpha ({\mathbf{x}} , t, \varepsilon_\alpha) $. It means that $%
\displaystyle \delta_\alpha a\ = \frac {da}{ d\varepsilon_\alpha}\
\vert_{\varepsilon_\alpha } = 0 . $ We have to emphasis on the fact
that the domain $ D\, $ is fixed in the physical space. This
particularity is related to the impossibility to have material
volume in general motion of the mixture. Taking into account
formulae (\ref{(displacement)}), variations in Appendix A and
definitions (\ref{(definitions)}), we get
\[
\delta_\alpha a = \int_\omega\ \bigg(R_\alpha \delta \rho_\alpha +
\rho_\alpha\, {\mathbf{K}}_\alpha^\star \delta{\mathbf{u}}_\alpha -
\rho_\alpha \theta_\alpha \delta s_\alpha\bigg) d\omega
\]
\[
=\int_{\omega_\alpha} \bigg(R_\alpha\, div_\alpha (\rho_{\alpha 0} \delta {%
\mathbf{X}}_\alpha) - \rho_{\alpha 0}\, {\mathbf{K}}^\star_\alpha\ F_\alpha\
\frac{\partial}{ \partial t} (\delta {\mathbf{X}}_\alpha) - \rho_{\alpha
0}\, \theta_\alpha\ \frac {\partial s_{\alpha 0}}{ \partial {\mathbf{X}}%
_\alpha}\ \delta{\mathbf{X}}_\alpha\bigg) d\omega_\alpha .
\]
In the last expression all quantities are considered as functions of $\ ({%
\mathbf{X}}_\alpha , t)$. Hence,
\[
\delta_\alpha a = \int_{\omega_\alpha} \rho_{\alpha 0} \left(
-\frac{\partial R_\alpha}{ \partial{\mathbf{X}}_\alpha} + \frac {\partial}{%
\partial t} ({\mathbf{K}}_\alpha^\star\ F_\alpha)\ - \ \theta_\alpha\frac {%
\partial s_{\alpha 0}}{ \partial{\mathbf{X}}_\alpha} \right)
\delta{\mathbf{X}}_\alpha\ d\omega_\alpha
\]
\[
+ \int_{\omega_\alpha}
Div_\alpha (\rho_{\alpha 0}\ {\mathbf{G}})\ d\omega_\alpha\ = 0
\]
where $\ {\mathbf{G}} = (R_\alpha\ \delta{\mathbf{X}}_\alpha ,\, - {\mathbf{K}}%
_\alpha^\star\ F_\alpha\ \delta{\mathbf{X}}_\alpha)\ $ and $\ Div_\alpha\ $
is the divergence operator in the 4-dimensional space $\,{\omega_\alpha}\,$
associated with $({\mathbf{X}}_\alpha , t)$. All the functions are assumed
to be smooth enough in the domain $\, \omega_\alpha\, $ and $\ \delta{%
\mathbf{X}}_\alpha \ = \ 0\ $ on $\ \partial \omega_\alpha. $ Then, we
obtain the equations of motion for each component in Lagrangian coordinates:
\begin{equation}
\frac {\partial}{\partial t}\ ({\mathbf{K}}^\star_\alpha\ F_\alpha)\ -\frac {%
\partial R_\alpha}{ \partial{\mathbf{X}}_\alpha}\ - \theta_\alpha\,\frac {%
\partial s_{\alpha 0}}{ \partial{\mathbf{X}}_\alpha}\ = 0  \label{(equimage)}
\end{equation}
Taking into account the identity $\displaystyle\
{\frac{d_\alpha F_\alpha}{dt}} \,-\, {\frac{\partial{\mathbf{u}}_\alpha}{%
\partial{\mathbf{x}}}}\ F_\alpha = 0, $ we rewrite (\ref{(equimage)}) in
Eulerian coordinates in the form :
\begin{equation}
\frac{d_\alpha{\mathbf{K}}_\alpha^\star}{ dt}\ +\, {\mathbf{K}}%
_\alpha^\star\ \frac {\partial{\mathbf{u}}_\alpha}{ \partial{\mathbf{x}}}
\,=\,\frac {\partial R_\alpha}{ \partial{\mathbf{x}}}\ + \theta_\alpha\frac {%
\partial s_\alpha}{ \partial{\mathbf{x}}}
\label{(motions)}
\end{equation}
\noindent If $\ {\mathbf{u}}_1 = {\mathbf{u}}_2 = {\mathbf{u}}\,,$ then $\ {%
\mathbf{K}}_\alpha\ =\ {\mathbf{u_{}}}\ $ and  (\ref{(equimage)}) is
equivalent to
\[
{\frac{d{\mathbf{u}}}{dt}}\ +\ \nabla^\star\, (h \, +\, \Omega)\ = \ \theta
\ \nabla^\star s\ ,\ \ \ \ {\frac{d}{dt}}\ = \ {\frac{\partial}{\partial t}}%
\ + \ {\mathbf{u}}^\star\ \nabla^\star
\]
\noindent where $h $ is the enthalpy and $\Omega\ $ is an external
potential \cite{Serrin}. Conservations of the total momentum and the
total energy are a consequence of the governing equations
(\ref{(mass)}), (\ref{(motions)}):
\begin{equation}
\sum^2_{\alpha=1} \frac {\partial
{\rho_\alpha\,\mathbf{K}}^\star_\alpha }{
\partial t}\, +\, div \bigg(\rho_\alpha\,{\mathbf{u}}_\alpha\,{\mathbf{K}}%
_\alpha^\star + \Big(\rho_\alpha\, \frac{\partial W }{ \partial\rho_\alpha}
- \ W\Big)\ I \bigg)\, + \rho_\alpha\, \frac{\partial \Omega_\alpha }{
\partial{\mathbf{x}}}  = 0  \label{(totalmomentum)}
\end{equation}
\begin{equation}
\sum^2_{\alpha=1} \frac {\partial}{ \partial t} \left( \rho_\alpha
\left( \frac{1}{2} {\mathbf{u}}_\alpha^2 + \Omega_\alpha \right)+
U\right) +div \Big( \rho_\alpha {\mathbf{u}}_\alpha\,
\left({\mathbf{K}}_\alpha^\star\ {
\mathbf{u}}_\alpha - R_\alpha \right)\Big) -\rho_\alpha \frac{%
\partial\Omega_\alpha}{ \partial t}=0  \label{(totalenergy)}
\end{equation}
The covector $\, {\mathbf{K}}^\star_\alpha\, $ is an essential
quantity; indeed, $\rho_\alpha {\mathbf{K}}_\alpha\, $ (but not $\,
\rho_\alpha{\mathbf{u}}_\alpha\, $) is the impulse for the $\,
\alpha^{th}$ component of the mixture. Morever, for adiabatic
motions, the
definition of potential flows for two-component mixtures is associated with $%
\ rot \ {\mathbf{K}}_\alpha = 0\ $  and not with \ $rot \ {\mathbf{u}}%
_\alpha = 0$. For potential motion, equation (\ref{(motions)})
yields additional conservation laws \cite{Gavrilyuk2,Gavrilyuk4}:
\[
\frac{\partial{\mathbf{K}}_\alpha^\star}{ \partial t}\, + \, \nabla\ ({%
\mathbf{K}}^\star_\alpha \ {\mathbf{u}}_\alpha - R_\alpha)\ =\ 0
\]
\noindent In the particular case of bubbly liquids, Geurst was the
first to carry out a term analog to ${\mathbf{K}}_\alpha$
\cite{Geurst}.  Moreover, if $\mathrm{rot} \, {\mathbf{K}}_\alpha\-
\ne {0}, \, $ the system of governing equations is not conservative
in terms of $\, {\mathbf{K}}_\alpha , \, \rho_\alpha\ $ and $\,
s_\alpha\, $ (the number of conservation laws admitted by the system
is less than the number of unknown variables); but, nevertheless the
system can be rewritten in conservative form if we add the gradient
tensor $\, F_\alpha\, $ as unknown variable \cite{Gavrilyuk2}.

\section{Governing equations in the dissipative case and dynamic Gibbs
identity}

The conservative fluid mixture model presented in section 2 is
relevant to the \textit{first gradient theory} \cite{Germain}: the
forces applied to the continuous medium are divided into volume
forces and surface  forces. In fluid mixture flows it is reasonable
to neglect the surface friction forces compared  to galilean
invariant algebraic volume forces.  The virtual work $\, \delta
\mathcal{T}_\alpha\, $ of dissipative forces applied to the
$\alpha^{th}$
component is in the form $\, \displaystyle \delta \mathcal{T}_\alpha\ = \ {%
\mathbf{f}}^\star_\alpha\ \delta_\alpha {\mathbf{x}}$.   For the
same virtual displacement of two components, $\, \delta{ \mathbf{x}}
= \delta_\alpha {\mathbf{x}}_1 = \delta_\alpha {\mathbf{x}}_2 \, $,
the total virtual work of dissipative forces is $\displaystyle
\delta\mathcal{T}\ = \ \sum^2_{\alpha=1}\ {\mathbf{f}}^\star_\alpha\ \delta{%
\mathbf{x}}$.\\  For a solid displacement, the work $%
\delta\mathcal{T}\, $ is equal to zero and consequently $\displaystyle %
\sum^2_{\alpha=1}  {\mathbf{f}}^\star_\alpha=0. \ $  We specify
later the behavior of forces $\, {\mathbf{f}}^\star_\alpha\, $. Let us introduce the quantities $\, {\mathbf{M}}%
_\alpha, \, B_\alpha ,\, S\, $ and $E\, $ such that:\newline

$\displaystyle {\mathbf{M}}^\star_\alpha\ = \ \rho_\alpha\frac {d_\alpha{%
\mathbf{K}}^\star_\alpha}{ dt}\ +\ \rho_\alpha\ {\mathbf{K}}^\star_\alpha%
\frac {\partial {\mathbf{u}}_\alpha}{ \partial{\mathbf{x}}}\ -\ \rho_\alpha%
\frac {\partial R_\alpha}{ \partial{\mathbf{x}}}\ - \rho_\alpha\
\theta_\alpha\frac {\partial s_\alpha}{ \partial{\mathbf{x}}} - \
{\mathbf{f}^\star_\alpha}  $

$\displaystyle {B}_\alpha \ = \frac {\partial\rho_\alpha}{ \partial
t}\ + \ div (\rho_\alpha{\mathbf{u}}_\alpha)$

$\displaystyle S\ = \ \sum^2_{\alpha = 1}\   \rho_\alpha\
\theta_\alpha\frac {d_\alpha s_\alpha}{ dt}\ +\ {\mathbf{f}}^\star_\alpha\ {%
\mathbf{u}}_\alpha\  $

$\displaystyle E = \sum^2_{\alpha=1}\, \frac { \partial }{\partial t}\, \bigg( %
\rho_\alpha\, (\frac{1}{2} \, {\mathbf{u}}_\alpha^2 \, +
\Omega_\alpha)  + U\bigg) \ +  div \bigg(
\rho_\alpha\ {\mathbf{u}}_\alpha\ ({\mathbf{K}}^\star_\alpha\ {\mathbf{u}}%
_\alpha \ -\ R_\alpha)  \bigg)  - \rho_\alpha\,\frac {\partial
\Omega_\alpha}{ \partial t}  $
\newline  We prove in Appendix $\, B \, $ the
following property:\newline

\noindent\textbf{Theorem:}\textit{\ \ For any motion of the mixture,
we have the identity}
\[
E \ -\ \sum^2_{\alpha=1}\ \bigg(M^\star_\alpha\ {\mathbf{u}}_\alpha \ +\ ({%
\mathbf{K}}^\star_\alpha\ {\mathbf{u}}_\alpha \ -\ R_\alpha)\ B_\alpha\bigg) %
\ -\ S\ \equiv 0
\]
This relation is the most general expression of the \textit{Gibbs
identity} \ in dynamics. Analogous identities were obtained earlier
for thermocapillary mixtures \cite{Gouin} and bubbly liquids
\cite{Gavrilyuk3}. For each component of the mixture, equation of
momentum and equation of mass are in the form
\begin{equation}
{\mathbf{M}}_\alpha^\star \ =\ 0, \ B_\alpha \ =\ 0
\label{(massmomentum)}
\end{equation}
The Gibbs identity implies  $\, S \, = \, E $. Hence, the equation
of the entropy $S\, = \,0$,
\begin{equation}
\sum^2_{\alpha=1}\,   \rho_\alpha\, \theta_\alpha\, \frac{d_\alpha
s_\alpha}{dt} \ +\ {\mathbf{f}}^\star_\alpha\, {\mathbf{u}}_\alpha
=\, 0  \label{(planck)}
\end{equation}
is equivalent to the equation of the energy $E \, = \, 0$. We note
also that the equations \, ${\mathbf{M}}_\alpha^\star \, =\, 0$ and
\ $B_\alpha \, =\, 0\, $  imply conservation of the total momentum
of the mixture ${\mathbf{M}} \, = \, 0$ (see (\ref
{(totalmomentum)}))

\section{The second law of thermodynamics}

In conservative case, the system with two different entropies is
closed by (\ref{(mass)}). In dissipative case we need additional
arguments to obtain equations for each entropy $s_\alpha$ which
could replace equations (\ref{(mass)}). We take these equations in
the form :
\begin{equation}
\rho_\alpha\, \theta_\alpha\, \frac{d_\alpha s_\alpha}{dt} \ +\ {\mathbf{f}%
_\alpha}^\star\, ({\mathbf{u}}_\alpha -{\mathbf{u}}) + q_\alpha = 0
\label{(planckA)}
\end{equation}
 which must be compatible with (\ref{(planck)}).
Here $\displaystyle \rho\, {\mathbf{u}} = \sum^2_{\alpha=1}\,\rho_\alpha {%
\mathbf{u}}_\alpha$ is the total momentum, $\displaystyle \rho \sum^2_{\alpha=1}\,\rho_\alpha$ and
$\displaystyle\,\sum^2_{\alpha=1}\, q_\alpha = 0$. The last relation
means that we have only internal heat exchanges between components.
Consequently, if
\begin{equation}
\sum^2_{\alpha=1}\,
\,\frac{{\mathbf{f}}^\star_\alpha}{\theta_\alpha}\, ({
\mathbf{u}}_\alpha -\mathbf{u}) +\frac{q_\alpha}{\theta_\alpha}\,
\leq\, 0 , \label{(dissipation)}
\end{equation}
we obtain the entropy inequality \cite{Muller,Muller2}
\[
\sum^2_{\alpha=1}\, \rho_\alpha\, \frac{d_\alpha s_\alpha}{dt} \, \geq\, 0
\]
Let us note that relation (\ref{(dissipation)}) is verified if
\[
{\mathbf{f}}_1 \ =\ k\ (\frac{{\mathbf{u}}_2-{\mathbf{u}}}{\theta_2} \ -\ \frac{{\mathbf{u%
}}_1-{\mathbf{u}}}{\theta_1}) \, , \ \ {\mathbf{f}}_2 -{\mathbf{f}}_1 \ , \ k\
>\ 0
\]
and
\[
q_1= \kappa\ (\frac{1}{\theta_2}-\frac{1}{\theta_1}%
),\ \ q_1 = -q_2,  \  \ \kappa > 0
\]
 The fact that the inverse temperatures
(coldness) appear in the closure relations play an important role in
other applications \cite{Ruggeri}.

\section{Fick's law as a consequence of the governing equations}

The governing equations for each component are :
\[
{\mathbf{M}}^\star_\alpha\, \equiv\, \rho_\alpha\, \frac{ d_\alpha{\mathbf{K}%
}_\alpha^\star}{dt} \, +\, \rho_\alpha\, {\mathbf{K}}^\star_\alpha\, \frac{%
\partial {\mathbf{u}}_\alpha}{ \partial{\mathbf{x }}} \, -\, \rho_\alpha\,
\frac{\partial R_\alpha }{ \partial {\mathbf{x}}_\alpha} \, -\,
\rho_\alpha\, \theta_\alpha\, \frac{\partial s_\alpha}{ \partial {\mathbf{x}}%
} \,  -\ {\mathbf{f}}_\alpha^\star \, =\, 0
\]
For  slow isothermal motions ($\theta_1=\theta_2=\theta_0=const$),
we can rewrite these equations in the following approximate form:
\[
{\mathbf{M}}^\star_\alpha\, \simeq\, \rho_\alpha\, \frac{\partial }{
\partial {\mathbf{x}}}\,\frac {\partial W}{\partial \rho_\alpha} \, -\,
\rho_\alpha\, \theta_0\, \frac{\partial s_\alpha}{ \partial {\mathbf{x}}%
} \,  -\, {\mathbf{f}}^\star_\alpha \, =\, 0
\]
or
\[
{\mathbf{M}}^\star_\alpha\, \simeq\, \rho_\alpha\,\frac {\partial
\mu_\alpha }{ \partial {\mathbf{x}}}\, -\, {\mathbf{f}}^\star_\alpha
\, =\, 0
\]
where $\displaystyle \mu_\alpha = \frac {\partial W}{ \partial \rho_\alpha}
- \theta_0\ s_\alpha \ $ is the chemical potential for the $\ \alpha^{th}$
phase. Considering the difference $\ {%
\mathbf{M}}^\star_2\ -\ {\mathbf{M}}^\star_1\ $ we obtain :
\begin{equation}
\nabla \mu \, = \, \frac {{\mathbf{f}}_2^\star }{ \rho_2} - \frac{{\mathbf{f}%
}^\star _1}{ \rho_1}\, \equiv\, \frac {\rho \ {\mathbf{f}}^\star }{
\rho_1\rho_2} \label{(chemical)}
\end{equation}
where ${\mathbf{f}}^\star  = -\, {\mathbf{f}}%
_1^\star\,$  and $\, \mu \, =\, \mu_2 \, -\, \mu_1 $. Equation
(\ref{(chemical)}) is the general form of Fick's law. So, Fick's law
is not a linear phenomenological law  but a direct consequence of
equations of motion and the Stokes drag hypothesis which was
previously noticed by Bowen with an other model \cite{Bowen}.

\section{Hyperbolicity of the two-fluid mixture model}

The hyperbolicity of governing equations is very important because
it implies wellposedness of the Cauchy problem. In mechanical case
(when we neglect the equations of entropies) we are back to our
previous study \cite{Gavrilyuk2,Gavrilyuk4}. The only difference
is the right-hand side algebraic terms $\mathbf{f}_\alpha$ due to
the Stokes-like drag forces. Obviously, they do not affect the
hyperbolicity analysis.  The potential $W$ is then a function of
$\rho_1, \rho_2$ and $w= |{\mathbf{w}}|$. The Lagrangian is (with
$\Omega_\alpha=0$)$$ L = \sum^2_{\alpha=1} \ \frac{1}{2} \
\rho_{\alpha} \, \mathbf{u}_{\alpha}^2 - W \ (\rho_1\ , \rho_2\ ,
w)$$ \newline We gave a sufficient condition  of the hyperbolicity
of system (\ref{(mass)}), (\ref{(totalmomentum)}) in the
multi-dimensional irrotational case where \ $rot \
\mathbf{K}_{\alpha} \ = \ 0$. Recall the main results we obtained
in this case: after a change of variables, system
\big((\ref{(mass)}),(\ref{(totalmomentum)})\big) takes the form
\begin{equation}
{\frac{\partial }{\partial t}}\bigg( \ {\frac{\partial G}{\partial
\sigma_{\alpha}}} \ \bigg) \ - \ div \ \bigg( \ {\frac{\partial }{\partial
\sigma_{\alpha}}}\ \bigg( \ \sum^2_{\beta=1} \ \sigma_\beta \ \mathbf{j}%
_{\beta} \ \bigg)\ \bigg) \ = \ 0,  \label{(Neweq1)}
\end{equation}
\begin{equation}
{\frac{\partial }{\partial t}}\bigg( \ {\frac{\partial G}{\partial \mathbf{j}%
_{\alpha}}} \ \bigg) \ - \ div \ \bigg( \ {\frac{\partial }{\partial \mathbf{%
j}_{\alpha}}}\ \bigg( \ \sum^2_{\beta=1} \ \sigma_{\beta } \ \mathbf{j}%
_{\beta} \ \bigg)\ \bigg) \ = \ 0.  \label{(Neweq2)}
\end{equation}
where
\[
G (\sigma_{1},\ \sigma_{2}, \ \mathbf{j}_{1}, \ \mathbf{j}_{2}) \ \ L (\rho_{1}, \ \rho_{2},\ \mathbf{j}_{1}, \ \mathbf{j}_{2}) \ - \
\sum^2_{\alpha=1}\ \sigma_{\alpha}\ \rho_{\alpha},\    \ {\rm with}\
\ \sigma_{\alpha}=  {\frac{\partial L}{\partial\rho_{\alpha}}}
\]
The function \ $G$ \ is a partial Legendre transformation of \ $L\
(\rho_{1}, \ \rho_{2},\ \mathbf{j}_{1}, \ \mathbf{j}_{2})$ \ with respect to
the variables \ $\rho_{\alpha}$:
\[
{\frac{\partial G}{\partial \sigma_{\alpha}}} \ = \ - \ \rho_{\alpha}, \ \ \
\ {\frac{\partial G}{\partial \mathbf{j}_{\alpha}}} \ = \ \mathbf{K}%
^*_{\alpha},
\]
and the system (\ref{(Neweq1)}), (\ref{(Neweq2)}) can be rewritten
in a symmetric form \ \cite{Godunov61,Friedrichs71,Boillat}
\begin{equation}
A \ {\frac{\partial \mathbf{u}}{\partial t}} \ + \ B^i \
{\frac{\partial \mathbf{u}}{\partial x^i}} \ = \ 0 \,, \ \ A = A^*
,\ B^i \ = (B^i)^*, \ \ \ i = 1,\, 2,\, 3   \label{(Symmetric)}
\end{equation}
where
\[
\mathbf{u}^* \ = \ (\sigma_1 \, , \ \sigma_2 \, ,\ \mathbf{j}^*_1,\ \mathbf{j}%
^*_2) \ , \hskip 0,5 cm A \ = {\frac{\partial^2 G}{\partial \mathbf{u}^2}}
\]
and the matrices \ $B^i\,$   can be obtained from (\ref{(Neweq1)}), (\ref%
{(Neweq2)}). If \ $A$ \ is positive definite, system (\ref{(Symmetric)}%
) is hyperbolic. We proved in \cite{Gavrilyuk2,Gavrilyuk4} that the
conditions
\begin{equation}
{\frac{\partial^2 W}{\partial w^2}}< \ 0 \, , \ {\frac{\partial^2 W}{%
\partial\rho_1^2}}\ > \ 0, \ {\frac{\partial^2 W}{\partial\rho_1^2}}\ {\frac{%
\partial^2 W}{\partial\rho_2^2}}-\bigg( {\frac{\partial^2 W}{%
\partial\rho_1\partial\rho_2}} \bigg)^2 \ >0  \label{(inequalities)}
\end{equation}
guarantee the hyperbolicity of our system for small relative velocity of
phases. Due to relation (\ref{(internalenergy)}), the inequalities (\ref%
{(inequalities)}) mean the convexity of the internal energy  $ U\, $
that corresponds to a natural condition of stability. Finally, we
established that the  stability implies the hyperbolicity of the
governing equations for small relative velocity \ $\mathbf{w}$,
provided that \ $rot \ \mathbf{K}_\alpha = 0$ (condition always
fulfilled for  one-dimensional flows).

\section*{Acknowledgments}

We are grateful to Professor Tommaso Ruggeri for helpful
discussions.

\bigskip \centerline{{\bf Appendix A}.} \bigskip \noindent
The definition of
the Lagrangian coordinates $\ {\mathbf{X}}_\alpha\ $ implies\ \ $%
\displaystyle \frac {\partial {\mathbf{X}}_\alpha }{ \partial t}\,
 +\, \frac{%
\partial {\mathbf{X}}_\alpha}{ \partial {\mathbf{x}}}\
{\mathbf{u}}_\alpha \, =\, 0$.  Taking the derivative with respect
to $\ \varepsilon_\alpha\ $ at zero, we obtain the following
equation :
\[
\frac {\partial \delta {\mathbf{X}}_\alpha}{ \partial t} \ +\ \frac{\partial
\delta {\mathbf{X}}_\alpha }{ \partial {\mathbf{x}}} \ {\mathbf{u}}_\alpha
\, +\,\frac {\partial {{\mathbf{X}}_{\alpha}}}{ \partial {\mathbf{x}}}\
\delta {\mathbf{u}}_\alpha \ =\ 0 ,\ \  \delta {\mathbf{u}}_\alpha\, ({%
\mathbf{x}} , \, t) \, =\, -\ F_\alpha\,\frac {d_\alpha}{ dt}\ (\delta {%
\mathbf{X}}_\alpha)
\]
Equation (\ref{(determinant)}) yields :
\begin{equation}
\delta\rho_\alpha\, ({\mathbf{x}}, t)\, \det\, F_\alpha\,
({\mathbf{x}} , t) \, +\, \rho_\alpha\, \delta\, (\det\, F_\alpha)
\, =\, \frac{\partial \rho_{\alpha 0}}{\partial
{\mathbf{X}}_\alpha}\, \delta{\mathbf{X}}_\alpha
\label{(transform)}
\end{equation}
Using the Euler-Jacobi identity, $\ \displaystyle
\delta (\det\ F_\alpha ) \, =\, \det\, F_\alpha\, ({\mathbf{x}}, t)\ \ tr\ %
\bigg(F^{-1}_\alpha\ \delta F_\alpha\ \bigg) $ \newline and\ \
$\displaystyle \delta F_\alpha\ ({\mathbf{x}}, t) \, =\, -\
F_\alpha\
({\mathbf{x}} , t)\ \delta F^{-1}\ ({\mathbf{x}} , t)\ F_\alpha\ ({\mathbf{x}%
} ,  t)\, ,\ \ \ \delta F^{-1}\, ( {\mathbf{x}}, t) \,  = \,\frac
{\partial \delta{\mathbf{X}}_\alpha}{ \partial {\mathbf{x}}}\, $,\\
we deduce :
\[
\delta\ (\det \ F_\alpha\ )({\mathbf{x }} , t) \ =\ -\ \det \ F_\alpha\ \
tr\ \bigg(\delta F^{-1} \ F_\alpha\bigg)  \ =\ -\det\ F_{\alpha}\ tr\bigg(\,%
\frac {\partial{\ \delta {\mathbf{X}_\alpha}}}{ \partial {\mathbf{X}}_\alpha}%
\bigg)\ .\ \ \mathrm{Or,}
\]
\begin{equation}
\delta\det F_\alpha \ =\ -\det \, F_{\alpha}\ \ div_\alpha(\delta{\mathbf{X}}%
_\alpha)    \label{(legendre2)}
\end{equation}
Substituting (\ref{(legendre2)}) into (\ref{(transform)})  we obtain
:
\[
\delta\rho_\alpha({\mathbf{x}} , t) \ =\ \rho_\alpha\ div_\alpha(\delta {%
\mathbf{X}}_\alpha) \, +\, \, \frac{\rho_\alpha}{ \rho_{\alpha 0}}\ \frac{%
\partial \rho_{\alpha 0}}{ \partial {\mathbf{X}}_\alpha}\ \delta{\mathbf{X}}%
_\alpha = \frac{div_\alpha(\rho_{\alpha 0}\, \delta{\mathbf{X}}
_\alpha)}{det F_\alpha}
\]
\[
\delta s_\alpha\, ({\mathbf{x}} , t)\ = \frac {\partial s_{\alpha 0}}{%
\partial{\mathbf{X}}_\alpha}\ \delta {\mathbf{X}}_{\alpha}
\]
\pagebreak

\centerline{{\bf Appendix B}.}

\medskip

\noindent The proof of the Gibbs identity is obtained by summing the
following algebraic identities $\ a  - f : \ $

\noindent For dissipative  terms,

\noindent \textbf{a. }
\[
 {\mathbf{%
f}}_1^\star \ {\mathbf{u}}_1 \ +\ {\mathbf{f}}_2^\star \
{\mathbf{u}}_2 -\ {\mathbf{f}}^\star _1\ {\mathbf{u}}_1 \ -\ {\mathbf{f}}^\star _2\ {%
\mathbf{u}}_2  \equiv\ 0
\]

\noindent For the external potentials $\ \Omega_\alpha $,

\noindent \textbf{b.}
\[
\displaystyle  \ {\frac{\partial }{\partial t}}\ \rho_\alpha\Omega_\alpha \
+\ div\ (\rho_\alpha\Omega_\alpha{\mathbf{u}}_\alpha) \ -\ \rho_\alpha\ {%
\frac{\partial \Omega_\alpha}{\partial {\mathbf{x}}}}\ {\mathbf{u}}_\alpha \
-\ B_\alpha\Omega_\alpha \ -\ \rho_\alpha\ {\frac{\partial \Omega_\alpha}{%
\partial t}}\ \equiv\ 0
\]
\noindent For the velocity fields $\ {\mathbf{u}}_\alpha,$

\noindent \textbf{c.}
\[
\displaystyle  \ {\frac{\partial }{\partial t}}\ \left(\
{\frac{1}{2}}\
\rho_\alpha\  {\mathbf{u}}_\alpha^2 \right) \ +\ div\ \left(%
\rho_\alpha\ {\mathbf{u}}_\alpha\ ( {\mathbf{u}}_\alpha^2\ -\ {%
\frac{1}{2}}\  {\mathbf{u}}_\alpha^2)\right)\
\]
\[
\displaystyle  -\ B_\alpha\ \left( {\mathbf{u}}_\alpha^2\ -\ {%
\frac{1}{2}}\ {\mathbf{u}}_\alpha^2 \right) \ -\ \left(%
\rho_\alpha\ {\frac{d_\alpha{\mathbf{u}}^\star _\alpha}{dt}} \ +\
\rho_\alpha\ {\mathbf{u}}^\star_\alpha\ {\frac{\partial {\mathbf{u}}_\alpha}{%
\partial {\mathbf{x}}}}\ \ -\ \rho_\alpha\ {\frac{\partial }{\partial{%
\mathbf{x}} }}\ (\ {\frac{1}{2}}\ {\mathbf{u}}_\alpha^2) \right)%
\ {\mathbf{u}}_\alpha\ \equiv\ 0
\]
Let us introduce $\displaystyle \ {\mathbf{i}}%
^\star \ =\ -\ {\frac{\partial W}{\partial {\mathbf{w}}}}$ . Then the
expression,

\[
\displaystyle  \ {\frac{\partial U}{\partial t}}\ \equiv\ {\frac{\partial }{%
\partial t}}\ \bigg(W \ -\ \ {\frac{\partial W}{\partial {\mathbf{w}}}}\ {%
\mathbf{w}}\bigg) \ =\ {\frac{\partial \ {\mathbf{i}}^\star }{\partial t}}\ {%
\mathbf{w}} \ +\ \sum^2_{\alpha=1}\ \bigg(\ {\frac{\partial W }{\partial
\rho_\alpha}}\ {\frac{\partial \rho_\alpha}{\partial t}} \ +\ \rho_\alpha\
\theta_\alpha\ {\frac{\partial s_\alpha}{\partial t}}\ \bigg)
\]
\noindent and the three following identities $\ d - f\ $ prove the formula.

\noindent \textbf{d.}
\[
\displaystyle  \sum^2_{\alpha=1}\ {\frac{\partial W}{\partial
\rho_\alpha}}\ {\frac{\partial \rho_\alpha}{\partial t}}  + div
\left(
 {\frac{\partial W}{\partial \rho_\alpha}}\ \rho_\alpha\ {%
\mathbf{u}}_\alpha\right)
- \rho_\alpha\ {\frac{\partial }{\partial {\mathbf{x}}}}%
\ \bigg(\ {\frac{\partial W}{\partial \rho_\alpha}}\bigg)\ {\mathbf{u}}%
_\alpha  - {\frac{\partial W}{\partial \rho_\alpha}}\ %
\bigg({\frac{\partial \rho_\alpha}{\partial t}} \ +\ div\ (\rho_\alpha\ {%
\mathbf{u}}_\alpha)\bigg)\ \equiv \ 0
\]

\noindent \textbf{e.}
\[
\displaystyle   \sum^2_{\alpha=1}\ \rho_\alpha\ \theta_\alpha\ {\frac{%
\partial s_\alpha}{\partial t}} \ +\ \rho_\alpha\ \theta_\alpha\ {\frac{%
\partial s_\alpha}{\partial {\mathbf{x}}}}\ {\mathbf{u}}_\alpha \ -\
\rho_\alpha\  \theta_\alpha\ {\frac{d_\alpha s_\alpha}{dt}}\ \equiv \ 0
\]

\noindent \textbf{f.}
\[
{\frac{\partial {\mathbf{i}}^\star }{\partial t}}\ {\mathbf{w}} \
+\sum^2_{\alpha=1}\,  div \left(  (-1)^\alpha   \left(
{\frac{{\mathbf{i}}^\star }{ \rho_\alpha}} \, {\mathbf{u}}_\alpha
\right) \rho_\alpha {\mathbf{u}}_\alpha \right) - \left( \rho_\alpha
{\frac{d_\alpha}{dt}}  \left( (-1)^\alpha\,
{\frac{{\mathbf{i}}^\star }{\rho_\alpha}}\ \right)  +
\rho_\alpha (-1)^\alpha {\frac{{\mathbf{i}}^\star }{\rho_\alpha}}\, {\frac{%
\partial {\mathbf{u}}_\alpha}{\partial {\mathbf{x}}}}\right)\, {\mathbf{u}}%
_\alpha
\]
\[
\displaystyle  -\ (-1)^\alpha\ \left( {\frac{{\mathbf{i}}%
^\star }{\rho_\alpha}}\, {\mathbf{u}}_\alpha\right)\, \left(\
{\frac{\partial \rho_\alpha}{\partial t}}  + div\, (\rho_\alpha\
{\mathbf{u}}_\alpha) \right) \ \equiv\ 0
\]


\begin{thebibliography}{99}
\bibitem{Bowen} {Bowen, R.M., Theory of mixtures, in \textit{Continuum
physics}, Vol III, Ed. A.C. Eringen,  Acad. Press, London, 1976, pp.
1-127.}

\bibitem{Bedford} {Bedford, A., Drumheller, D.S., Recent advances. Theories
of immiscible and structured mixtures, \textit{Int. J. Engng. Sci.}, 1983,
\textbf{21}, 8, 863-960. }

\bibitem{Geurst} {Geurst, J. A., Variational principles and two-fluid
hydrodynamics of bubbly liquid/gas mixtures, \textit{Physica A}, 1986,
\textbf{135}, 455-486. }

\bibitem{Gouin} {Gouin, H., Variational theory of mixtures in continuum
mechanics, \textit{Eur. J. Mech, B/Fluids}, 1990, \textbf{9}, 469-491.}

\bibitem{Nigmatulin} {Nigmatulin, R. I., \textit{Fundamentals of mechanics
of heterogeneous mixtures}, Nauka (In Russian), Moscow,  1978. }

\bibitem{Saurel1} Gavrilyuk, S.L., Saurel, R., Mathematical and Numerical
Modeling of Two-phase Compressible Flows with Micro-Inertia,
\textit{J. Comp. Physics}, 2002, \textbf{175}, 326-360.

\bibitem{Gavrilyuk2} {Gavrilyuk, S. L., Gouin, H., Perepechko, Yu. V.,
Hyperbolic models of homogeneous two-fluid mixtures,
\textit{Meccanica}, 1998, \textbf{33}, 161-175.}

\bibitem{Muller} {M\"uller, I., \textit{Thermodynamics}, Interaction of
Mechanics and Mathematics Series, Pitman, London, 1985.}

\bibitem{Muller2} {M\"{u}ller, I., Ruggeri T., \textit{Rational Extended
Thermodynamics}, Springer, Berlin, 1998.}

\bibitem{Serrin} {Serrin, J., Mathematical principles of classical fluid
mechanics in: S. Fl\"ugge (Ed), \textit{Encyclopedia of Physics},
VIII/1, Springer, Berlin, 1959, pp. 125-263. }

\bibitem{Gavrilyuk4} {Gavrilyuk, S. L., Gouin, H., A new form of governing
equations of fluids arising from Hamilton's principle}, \textit{Int.
J. Engng. Sci.}, 1999, \textbf{37}, 1495-1520.



\bibitem{Germain} Germain, P., La m\'{e}thode des puissances virtuelles en m%
\'{e}canique des milieux continus,\emph{ Journal de M\'{e}canique}, 1973, \textbf{12%
}, 235-275.

\bibitem{Gavrilyuk3} {Gavrilyuk, S. L., Shugrin, S.M., Media with equations
of state that depends on derivatives, \textit{J. Appl. Mech. Techn. Physics}%
, 1996, \textbf{37}, 2, 177-189.}

\bibitem{Ruggeri} {Ruggeri, T., Relativistic extended thermodynamics in: A.
Anile, Y. Choquet-Bruhat (Eds), \textit{Relativistic Fluid
Dynamics}, Springer, Berlin, 1987.}

\bibitem{Godunov61} Godunov, S.K., An interesting class of quasilinear
systems, \textit{Sov. Math. Dokl.}, 1961, \textbf{2} , 947-949.

\bibitem{Friedrichs71} Friedrichs, K.O., Lax, P.D.,  Systems of conservation
laws with a convex extension, \textit{Proc. Nat. Acad. Sci. U.S.A.},
1971, \textbf{68} , 1686-1688.


\bibitem{Boillat} Boillat, G., Non-linear hyperbolic fields and waves in: T.
Ruggeri (Ed), \textit{Recent Mathematical Methods in Nonlinear Wave
Propagation}, Springer, Berlin, 1996, 1-47.
\end{thebibliography}
\end{document}